\def\conf{1}
\def\confcr{0}

\documentclass[11pt]{article}
\usepackage{fullpage}
\usepackage{graphicx,amsfonts,amsmath,amssymb,epsfig,hyperref,color}
\usepackage{multirow}
\usepackage{epsfig}

\renewcommand{\paragraph}[1]{{\protect\vspace{8pt}\noindent\sc{#1}}}

\ifnum\conf=0
\fi

\newlength{\saveparindent}
\newlength{\saveparskip}
\newcommand{\BE}{\begin{enumerate}} \newcommand{\EE}{\end{enumerate}}
\newcommand{\BI}{\begin{itemize}} \newcommand{\EI}{\end{itemize}}
\newcommand{\BDes}{\begin{description}}\newcommand{\EDes}{\end{description}}
\newtheorem{alg}{Algorithm}
\newcommand{\BA}{\begin{alg}} \newcommand{\EA}{\end{alg}}
\newtheorem{thm}{Theorem}[section]
\newcommand{\BT}{\begin{thm}} \newcommand{\ET}{\end{thm}}

\newtheorem{lem}{Lemma}[section]      % A counter for Lemmas
\newcommand{\BL}{\begin{lem}} \newcommand{\EL}{\end{lem}}

\newtheorem{clm}[lem]{Claim}
\newcommand{\BCM}{\begin{clm}} \newcommand{\ECM}{\end{clm}}

\newtheorem{techcor}[thm]{Corollary}
\newcommand{\BCo}{\begin{techcor}} \newcommand{\ECo}{\end{techcor}}

\newtheorem{cor}[thm]{Corollary}      % counter AS FOR Theorems
\newcommand{\BC}{\begin{cor}} \newcommand{\EC}{\end{cor}}
\newtheorem{prop}[thm]{Proposition}     % A counter AS FOR Thms
\newcommand{\BP}{\begin{prop}} \newcommand {\EP}{\end{prop}}
\newtheorem{conj} {Conjecture}      % counter AS FOR Theorems
\newcommand{\BCJ}{\begin{conj}} \newcommand{\ECJ}{\end{conj}}
\newtheorem{defn}{Definition}[section]         % A counter for Definition
\newcommand{\BD}{\begin{defn}} \newcommand{\ED}{\end{defn}}
\def\FullBox{\hbox{\vrule width 8pt height 8pt depth 0pt}}
\ifnum\confcr=0
\newcommand{\qed}{\;\;\;\FullBox}
\else
\renewcommand{\qed}{\;\;\;\FullBox}
\fi
\ifnum\confcr=0
\newenvironment{proof}{\noindent{\bf Proof:~~}}{\(\qed\)}
\fi
\newcommand{\BPF}{\begin{proof}} \newcommand {\EPF}{\end{proof}}
\newenvironment{proofof}[1]{\noindent{\bf Proof of {#1}:~~}}{\(\qed\)}
\newcommand{\BPFOF}{\begin{proofof}} \newcommand {\EPFOF}{\end{proofof}}
\newcommand{\qedsketch}{\;\;\;\Box}

\newenvironment{smallproof}{\noindent{\bf Proof:~~}}{\(\qedsketch\)}
\newcommand{\bpf}{\begin{smallproof}} \newcommand{\epf}{\end{smallproof}}
\newcommand{\BEQ}{\begin{equation}} \newcommand{\EEQ}{\end{equation}}
\newcommand{\BEQN}{\begin{eqnarray}}\newcommand{\EEQN}{\end{eqnarray}}
\newcommand{\eqdef}{\stackrel{\rm def}{=}}
\newcommand{\bitset}{\{0,1\}}

\renewcommand{\Pr}{{\rm Pr}}

\newcommand{\var}{{\mbox{\bf\rm Var}}}

\newcommand{\poly}{{\rm poly}}

\newcommand{\e}{\epsilon}
\newcommand{\eps}{\epsilon}

\newcommand{\todo}[1]{\iffalse {#1} \fi }

\newcommand{\Exp}{{\rm Exp}}

\newcommand{\tw}{{\widetilde{w}}}
\newcommand{\hw}{{\widehat{w}}}
\newcommand{\loglog}{{\rm loglog}}

\title{The Power of an Example: Hidden Set Size Approximation Using Group Queries and Conditional Sampling}
\author{
Dana Ron\thanks{
{\tt danaron@tau.ac.il}, Tel Aviv University.  Supported by ISF grant
number 671/13.}
\and
Gilad Tsur\thanks{
{\tt gilad.tsur@gmail.com}, Tel Aviv University.  Supported by
the Check Point Institute for Information Security .}
}

\begin{document}

\maketitle
\thispagestyle{empty}
\begin{abstract}
We study a basic problem of approximating the size of an unknown set $S$ in a known
universe $U$. We consider two versions of the problem. In both versions the algorithm
can specify subsets $T\subseteq U$. In the first version, which we refer to as
the {\em group query\/} or {\em subset query\/} version, the algorithm is told whether
$T\cap S$ is non-empty. In the second version, which we refer to as
the  {\em subset sampling\/} version, if $T\cap S$ is non-empty, then the algorithm
receives a uniformly selected element from $T\cap S$. We study the difference between these two versions under
different conditions on the subsets that the algorithm may query/sample, and
in both the case that the algorithm is adaptive and the case where it is non-adaptive. In particular we focus
on a natural family of allowed subsets, which correspond to intervals, as well as
variants of this family.
\end{abstract}

\newpage
\pagenumbering{arabic}

\section{Introduction}
Consider the following problem: For a known universe of elements $U$ and an unknown
 (``hidden'') subset of it $S\subseteq U$, we wish to approximate the size of $S$.
Of course, our ability to do so is influenced by what access we have to information about $S$. We consider two basic versions of this problem, where our algorithm has two different ways to obtain information about $S$:
\begin{itemize}
\item \textbf{Subset Queries} (also known as \textbf{Group Queries}): In this version we specify a subset $T \subseteq U$ and are told whether $S \cap T$ is empty or contains at least one element.
\item \textbf{Subset Samples}: In this, second, version we again specify a subset $T \subseteq U$, but are given a uniformly selected element of $S \cap T$ if such an element exists, and an indication that the intersection is empty otherwise.
\end{itemize}
Clearly, subset samples are at least as powerful as subset queries.  Note that subset queries (and hence subset samples) can be viewed as a generalization of membership queries (checking whether a single element in $U$ belongs to $S$). Also note that subset samples can be viewed as a generalization of sampling an element uniformly from $S$. We study the number of subset queries/samples required to approximate the size of $S$, where we may be restricted in which subsets $T\subseteq U$ we are allowed to query (or sample). This restriction is on the samples we may perform and not on the structure of $S$.

Stockmeyer~\cite{Stockmeyer83,Stockmeyer85}, in work better known for results on approximation
 algorithms for $\#{\cal P}$, considers the problem of set size approximation using what we call
subset queries.\footnote{We cite both the conference version~\cite{Stockmeyer83} and the journal version~\cite{Stockmeyer85} of Stockmeyer's paper, since some of the results appear only in the conference version.
 Subset queries are given different names in the two versions, including {\em intersection samples} and {\em subset samples}.} He considers the problem for different families of subsets that may be queried
and where the set $S$ may also be restricted to belong to a certain family.
% . In Stockmeyer's paper the set $S$ may also restricted to be from a certain family.
% Where technically relevant we mention these results, and the paper is further discussed (with other relevant results) in Subsection~\ref{subsec:related.research}.
We further discuss Stockmeyer's results (in particular in relation to our results) in Subsections~\ref{subsec:our-res} and~\ref{subsec:related.research}.
We mention briefly that subset queries arise in the context of {\em Group Testing\/}, where they are referred to as {\em group queries\/} (lending the paper part of its title). Estimation of the size of an unknown set appears in many different settings, including estimating the support size of a distribution and estimating the coverage of search engines. We discuss these settings as well in Subsection~\ref{subsec:related.research}.

%In one case all subsets may be queried. In the second case queries correspond to potential matchings in a graph. In the third case the subsets correspond to the vertices of a tree, and the elements of the universe to leaves. A sample contains members of the hidden set if such members are contained in the leaves of the subtree rooted in it.

%. Here the input (corresponding to the universe $U$ in our problem) is a circuit $C$, and we wish to approximate the number of values $x$ such that $C(x)=1$ (corresponding to the hidden set $S$). We are given a circuit satisfiability oracle, and it is used in a manner similar to our subset queries. The use of the oracle by Stockmeyer corresponds to specifying subsets of $U$ using small circuits.

\subsection{Precise Problem Definition}\label{subsec:prob-def}
We say that an algorithm is a {\em set size approximation algorithm\/} if,
for any set $S \subseteq U$, given
an approximation parameter $\eps\in (0,1]$ and either access to subset queries or to subset samples,
the algorithm returns an estimate $\hw$ such that with probability at least $2/3$,
$\frac{1}{1+\eps}|S| \leq \hw \leq (1+\eps)|S|$. The success probability of $2/3$ can be
increased to $1-\delta$ for any $\delta >0$  by standard techniques at a multiplicative cost of
$\log(1/\delta)$ in the complexity of the algorithms.
For an algorithm performing subset queries, the failure probability is over the
coin-tosses of the algorithm, and for an algorithm that uses subset samples, this
probability is also over the samples that the algorithm observes.
We are interested in the number of queries/samples used by the algorithm as a function
of the size of the universe $U$, which we denote by $n$, the the size of set $S$, which we denote by $w$, and the approximation
parameter $\eps$. This number may be
a random variable, in which case we shall bound the probability that it
exceeds $g(n,w,\eps)$ for a function $g$ that we shall specify.
We consider both {\em adaptive\/} algorithms and {\em non-adaptive\/} algorithms.
In this context, a non-adaptive algorithm cannot determine the next subset
it queries/samples based on previous answers to queries/samples, but it
can decide when to stop based on the information it obtains. This is necessary
in order to allow a dependence on the unknown value $w$ (rather than only
on $n$ and $\eps$).
% Since the complexity of the algorithm may depend on the unknown size $w$ of the set $S$,
% a non-adaptive algorithm must define a fixed distribution over sequences of
% subsets, and may stop and provide its output after obtaining answers to queries/samples
% for the subsets in prefix of the sequence.

\subsection{Our Results}\label{subsec:our-res}
We give upper and lower bounds for both versions of our problem and for both
non-adaptive and adaptive algorithms in several settings, as detailed next.
In what follows we focus on the dependence on $n$ and $w=|S|$. %\mnote{D: $|S|$ vs. $w$..}
In all cases the dependence on $1/\eps$ in the upper bounds is polynomial, and the
lower bound holds for constant $\eps$
(the exact dependencies appear in the statement of each theorem). The results are (generally) asymptotic and we use the $\tilde{O}(f)$ notation to hide a dependency on $\log(f)$. Thus, e.g., $\tilde{O}(\loglog(n))$ may hide a ${\rm logloglog}(n)$ factor.

%Our starting point
The simplest case we consider is the most basic and restricted setting in which no strict subsets of $U$ are allowed, namely,
the only subset allowed is the whole universe $U$. Obviously, with  subset queries it
 is not possible to approximate the set size (beyond distinguishing between an empty and non-empty set $S$).
 On the other hand, subset samples (which give us uniformly selected elements from $S$) can
 provide us with a good approximation when using a number of samples that grows like the square root of $w$ (and this dependence on $w$ can easily be seen to be necessary).
 As there is only one subset we can access, adaptivity plays no role in this setting.

In the other extreme we may allow the queries/samples to be on {\em all subsets\/}. Stockmeyer~\cite{Stockmeyer83} studied this problem for adaptive subset queries, and gave upper and lower bounds for this case.\footnote{Stockmeyer actually goes further and shows that a significantly smaller set of queries suffices to get similar results. He considers subsets of the elements in $U$ that are defined by a hash function on their index. This allows the family of subsets to be polynomial and not exponential in $n$.}
We refine these results to be in terms of $n,w$ and $\epsilon$, and not only in terms of $n$. As we are interested in the difference between the adaptive and non-adaptive setting we also give a non-adaptive version of these results.
%, and we consider the problem for non-adaptive subset queries and also give results for adaptive subset queries in terms of the precision $\epsilon$ and the size of the hidden set, where Stockmeyer gives results only for the size of the universe $n$.
In the non-adaptive version of this problem the number of queries has a logarithmic dependance on $w$, whereas in the adaptive case the dependence is
doubly logarithmic. These dependencies are tight in the case of subset queries.

%\item
%\mnote{D: I want to work more on this par}
Between the two extremes  we consider a natural family of subsets, which we refer to as
{\em Interval subsets\/}. In this case the universe $U$ is fully ordered, and the allowed subsets
 correspond to intervals of the elements. That is, subsets are of the form
 $T = \{u_i,u_{i+1},\dots,u_j\}$ where $U = \{u_1,u_2,\dots,u_n\}$.
Interval queries have been studied  in the context of group testing: Such a query may, e.g., be to see if a certain part of an electric circuit is faulty, or may be to a subset of ordered test-tubes that is more easily accessed by a robotic arm~\cite{DH}.
All of our positive results for interval subsets extend to the more general case of
grids in higher (constant) dimension $d$ where the allowed subsets correspond to $d$-dimensional
sub-grids,\footnote{In fact, the original motivation for this work arose from
our interest in estimating the number of 1-pixels in parts of a two-dimensional image.} and to the Boolean hypercube, where the allowed subsets correspond to sub-cubes.
%\mnote{D: double-check after writing for hypercube}

% As observed by Stockmeyer~\cite{Stockmeyer85},
Non-adaptive interval queries can be used to approximate the size of a set in a straightforward manner by using the fact that single elements in the universe are also intervals. This implies an upper bound of $O(n/w)$
on the (non-adaptive) query complexity, and we provide a simple matching lower bound. Hence,
non-adaptive interval queries do not offer a real advantage over just being able to query single elements of the universe selected uniformly at random. This situation changes when adaptivity is allowed. Specifically,
one of the results in~\cite{Stockmeyer85} contains (implicitly) an upper bound for adaptive interval queries of $O(\min(w\log(n),n/w))$. We give a lower bound of
$\Omega(\min\{w\cdot \log(n/w^2),n/w\})$, which shows that this is (almost) tight.\footnote{Stockmeyer~\cite{Stockmeyer85}  proves a lower bound for the related problem he studies,
but since the allowed query sets in his case are more restricted, the implications on interval queries
are not clear (furthermore, he proves his lower bound for sets of a particular size).}

% In the details of one of his proofs Stockmeyer uses a reduction to interval queries and gives bounds that correspond to our upper bounds for adaptive interval queries \mnote{stress that not samples}. These are given in terms of $n$ (without $w$ and $\epsilon$).
% The hypercube setting corresponds to the setting of trees studied by Stockmeyer, but here the more interesting results are for subset samples and not queries.

% Non-adaptive interval queries can be used to estimate the size of a set in a straightforward manner by using the fact that single elements in the universe are also intervals. This is used by Stockmeyer~\cite{Stockmeyer85}. Such a use of interval queries allows us to estimate $w$ using $n/w$ queries, and we give a lower bound that shows this is tight, that is, non-adaptive interval queries do not offer a real advantage over just being able to query single elements of the universe selected uniformly at random. One of the results in~\cite{Stockmeyer85} contains (implicitly) an upper bound for adaptive interval queries of $O(\min(n/w,w\log(n)))$, and we give a lower bound of $\Omega(\min\{w\cdot \log(n/w^2),n/w\})$ that shows this is (almost) tight.

Turning to interval samples (in contrast to queries) we get from the results mentioned above that
$O(\min\{n/w, \sqrt{w}\})$ such non-adaptive subset samples suffice to approximate the set size.
 Here too we give a matching lower bound.
 The power of samples when compared to queries really comes to fore when we present an adaptive  algorithm that uses
 $O({\rm poly}(\log(w)))$ interval subset samples. This algorithm can be adapted to additional settings such as $d$-dimensional grids and the Boolean hypercube, as we explain in Section~\ref{sec:int}.

%For Interval subsets there is an advantage to subset sampling over subset queries,
%but it is less dramatic than in the most restricted case.
%Specifically, for non-adaptive algorithms and $w < n^{2/3}$
%the complexity can be reduced from $O(n/w)$ to $O(\sqrt{w})$ and for adaptive
%algorithms it can be reduced from $O(w\log(n))$ to ${\rm \poly}(\log(w))$.
%It can also be seen that here too adaptivity improves the complexity,
%and in all but one case we prove that our bounds are (almost) tight.
% Here, while subset sampling certainly enjoys an advantage over group testing, the task can be accomplished by both. It is worth noting, however, that when sampling adaptively on intervals, in the {\GQ} setting we can do no better than identifying  each  of the elements separately, whereas in subset sampling we can perform the task exponentially faster.
% \item Unrestricted Subsets: Here we can sample (or query) any subset we wish. While lower bounding the number of {\SuS}s required seems somewhat harder than bounding the number of {\GQ}s,  we have no indication that they enjoy any speedup in contrast to group queries, and whatever speedup they do enjoy must be more limited than in the case of interval samples.

% \item Subsets of size $k$ - Here we can sample or query any subset of size exactly $k$. Where the three results above are ordered - interval sets are a strict refinement of $U$, and unrestricted subsets are a strict refinement of interval sets, this result isn't ????
% \mnote{D: Commented $k$ subsets. Mention in further research?}

% \EE

\begin{table}[ht]
\begin{center}
\begin{tabular}{|c|c c l|c c l|}
    \hline
    \textbf{Subsets allowed} & \multicolumn{3}{|c|}{\textbf{Subset Queries}} & \multicolumn{3}{|c|}{\textbf{Subset Sampling} }\\
%    \hline
%    None (Only whole universe) & \multicolumn{3}{|c|}{UB: $\infty$,LB: $\infty$} & \multicolumn{3}{|c|}{UB: $O(\sqrt{w})$,LB: $\Omega(\sqrt{w})$}\\
    \hline
    \multirow{2}{*}{Only $U$} & & UB: & $\infty$                                 & & UB: & $O(\sqrt{w})$\\
                                                & & LB: & $\infty$                                 & & LB: & $\Omega(\sqrt{w})$\\
    \hline
    \multirow{4}{*}{$\mbox{Interval}^{(*)}$}  & \multirow{2}{*}{NA:} & UB: & $O(n/w)$~\cite{Stockmeyer85}                                 & \multirow{2}{*}{NA:} & UB: & $[O(\min\{n/w, \sqrt{w}\})]$\\
                               &                      & LB: & $\Omega(n/w)$                            &                      & LB: & $\Omega(\min\{n/w, \sqrt{w}\})$\\
                               & \multirow{2}{*}{A:}  & UB: & $O(w\cdot \log(n))$~\cite{Stockmeyer85}                      & \multirow{2}{*}{A:}  & UB: & $O({\rm poly}(\log(w)))$\\
                               &                      & LB: & $\Omega(\min\{w\cdot \log(n/w^2),n/w\})$ &                      & LB: & \\
    \hline
    \multirow{4}{*}{All} % [Sto85]
    & \multirow{2}{*}{NA:} & UB: & $\tilde{O}(\log(w))$                     & \multirow{2}{*}{NA:} & UB: & $[\tilde{O}(\log(w))]$\\
                               &                      & LB: & $\tilde{\Omega}(\log(w))$                &                      & LB: & \\
                               & \multirow{2}{*}{A:}  & UB: & $\tilde{O}(\loglog(w))$~\cite{Stockmeyer83}                  & \multirow{2}{*}{A:}  & UB: & $[\tilde{O}(\loglog(w))]$\\
                               &                      & LB: &
                               $\tilde{\Omega}(\loglog(w))$~\cite{Stockmeyer83}    &                      & LB: & \\
    \hline
\end{tabular}
\end{center}
\caption{\small{A table of results. `NA' stands for Nonadaptive and `A' for Adaptive.
`UB' stands for Upper-bound and `LB' for Lower-bound. The dependence on $1/\eps$ in all upper
bounds is polynomial. We put brackets around those results that follows directly from
other cells in the table, and do not explicitly indicate the $O(n/w)$ upper bound that
holds in almost all cases. The results by Stockmeyer are originally as a function of $n$.
$(*)$ The upper bounds
% and some of the lower bounds
for the case of intervals apply to $d$-dimensional grids where $d$ is a constant and to the Boolean hypercube.} }\label{table.results.summary}
\end{table}
The results  described above are summarized in Figure~\ref{table.results.summary}.
As the results in each row of the table are for refinements of the row above it, one can always use upper bounds from cells above and to the left and lower bounds from cells below and to the right. Obviously  lower bounds for adaptive algorithms apply to non-adaptive ones and non-adaptive upper bounds are true also for adaptive algorithms.
% Furthermore, since the answers to subset queries are binary
% valued, the number of queries performed by a non-adaptive algorithm is at most exponential
% in the number of queries performed by an adaptive algorithm.

\subsection{Discussion and Techniques}
%\mnote{D: want to change this subsection - more focus on our main points}
When reviewing the results described in the previous subsection, the following relationship can be gleaned: As our ability  to specify subsets becomes less restricted, the advantage  of subset samples over subset queries (i.e., of getting a uniformly selected member of the subset) decreases. An interesting problem is to formalize and prove such a relationship. We conjecture that subset samples do not give any significant speedup when the subsets we have access to are unrestricted. % Proving this remains an open problem as well.
Another general phenomenon (which is of course not unique to the problems we consider), is the power
of adaptivity, which comes into play both when using subset queries and when using subset samples.

%%%%%%%%%%%%%%%%%%%%%%%%%%%%%%%%%
\iffalse
The  algorithms we describe are generally quite simple. In particular they rely on hitting
probabilities in the case of subset queries, which are used to search for a
good approximation of the set size, and on collision probabilities in
the case of subset sampling. In the case of interval subsets and subset sampling
we use the samples in a more interesting manner, as we elaborate
below. While the algorithms are in general simple, our lower bounds show
that in almost all cases, the complexity cannot be reduced by using more
sophisticated algorithms.

The lower bounds mostly apply the following approach, which is a variation
on standard lower bound techniques.
We give a distribution over pairs of instances, where in one
instance we ``hide'' a smaller subset and in the other we ``hide''
a larger subset. Any set size approximation algorithm should
be be able to distinguish between the two instances. We then lower
bound the number of subset queries an algorithm must perform in order to distinguish between
pairs chosen from this distribution. A more detailed discussion of this approach is
given in Section~\ref{sec:int}.
\fi
%%%%%%%%%%%%%%%%%%%%%%%%%%%%

Whereas some of the algorithms and lower bounds we present are fairly easy to establish, they
help to outline the relationship between subset queries and subset sampling.
In what follows we highlight two results. The first is the adaptive algorithm that uses interval subset sampling,
and which can be adapted to other families of subsets. The second is
the lower bound for adaptive interval subset queries. We believe that there are structural aspects
of the proof of this lower bound that are interesting.

In the adaptive interval sampling algorithm we maintain a sequence
 of intervals $I_0,\dots,I_t$, starting from an initial interval $I_0=U$,
 and ending with an interval containing a single element of $S$ (with high probability).
 The algorithm works iteratively, where in each iteration it continues with
 a subinterval $I_j$ of the previous interval $I_{j-1}$.
Each new subinterval $I_j$
 is selected (using sampling) so that the size of $I_j$'s intersection with $S$ is
 a constant fraction (not far from $1/2$) from the size of the intersection of
 $I_{j-1}$ with $S$. This ratio between the sizes of the two intersections is
 then estimated, and the final output is based on the sequence of these estimations.
% we sample some members of $S$ and divide the interval in a way that separates our sample
% in two. We estimate the relation between one side of the interval and the other and
% then proceed to recursively evaluate only one side of the interval.
% We show how this algorithm can be adapted to a variety of other subset families (e.g., grids with a fixed dimension
% $d$) and gives us a significant speed-up on the trivial approach.

In the lower bound for adaptive interval  queries, we ``force" any algorithm to deal separately with roughly $w$ different parts of the
universe and ``make it" use $\Omega(\log(n/w^2))$ queries for any such part. To formalize this we prove a type of direct-sum claim for a basic problem of determining the location
of a single `1' in a binary string. This allows us to come close to matching the upper bound.

% Moved this self critique to Future Research
%%%%
% The setting where we clearly lack techniques for lower bounds is that of adaptive
% subset sampling (as can be seen in Table~\ref{table.results.summary}). The only lower
% bound we have is for the setting where adaptivity is actually meaningless, as all the
% samples performed are on the whole universe.
\subsection{Related Research}\label{subsec:related.research}

As mentioned above, subset queries (in contrast to subset samples) were studied by Stockmeyer~\cite{Stockmeyer83, Stockmeyer85} in a different context, motivated by complexity theoretical problems (e.g., approximating the number of inputs that satisfy a circuit). Where in the current work we restrict the subsets that the algorithm may query but allow the set $S$ to be arbitrary, Stockmeyer restricts the set itself in addition to restricting the queries. He studies this problem for adaptive algorithms only, and gives  results for three families of subsets. The one that is analogous to a case in our work is when neither the subsets that are queried, nor the set $S$, are restricted (as mentioned above).  A second family is where the queries and the set must conform to a certain rooted tree configuration.
%, where if a vertex belongs to the subset queried (or to $S$) so must all of the path from this vertex to the root of the tree.
As mentioned previously, the algorithm given for this case translates directly to our adaptive interval query condition.
The third family of subsets is related to matchings in a graph, and does not appear to have direct relevance to our work. All the upper and lower bounds given by Stockmeyer, in contrast to those in this paper, are in terms of $n$ and not of $w$ and $\epsilon$.

Subset queries arise also in the context of {\em Group Testing\/}. In group testing, the algorithm is given access to subset queries (referred to as {\em group queries\/}) but the goal is to return the set $S$ itself rather than an estimate of the size of $S$. Research in the field of group testing began in the early 1940's \cite{Dorfman43} and is still of interest today (e.g. \cite{INR10}). Many different settings are considered in group testing, with a strong focus on non-adaptive vs. adaptive tests.
Similarly to what is done in this work, there is work on group testing when there are limitations on the subsets that the algorithm can query (for more details, see, e.g.,~\cite{DH}). The process we describe can be thought of as a relaxation of group testing, or may be used to get a quick estimate of the number of members in the hidden set, as a preliminary stage for Group Testing.
% SAY MORE ABOUT GROUP TESTING AND THAT WE GIVE RELAXATION/ PRELIMINARY STEP? REF K-COLOR TESTING PAPER (NOT CLEAR THAT REALLY RELEVANT)?

In the practical setting of estimating the relative coverage of search engines,
Bharat et~al.~\cite{BB98} consider the problem of
estimating the relative size of two sets by performing uniform queries on a set and then checking whether the elements sampled appear in the other set (and vice-versa). These operations, in our setting, are modeled by performing subset samples on the entire universe and by performing subset samples on single elements. Broder et al.~\cite{BFJKMNPTX06} tackle similar problems, and mention the great utility uniform sampling may have for solving them. Anagnostopoulos et~al.~\cite{ABC06} consider some additional problems related to sampling from search engines in an attempt to estimate sizes. None of the results correspond directly to ours.
% \mnote{D: non of the results are "same" as ours}

Approximating the size of a set using subset samples can be viewed as a special case of approximating the support size of a distribution using conditional sampling. When given access to samples generated according to a general distribution over a domain of size $n$ where each element in the support has probability at least $1/n$, there are almost linear lower bounds for approximating the support size~\cite{RRSS,Val,ValVal}.
The best lower bound, due to Valiant and Valiant~\cite{ValVal}, is $\Omega(n/\log(n))$, and this bound is tight~\cite{ValVal}. In the conditional sampling model~\cite{CFGM,CRS} it is possible to specify a subset $T$ of the domain of the distribution and obtain samples according
to the corresponding conditional distribution. The goal is to test various properties of the
distribution or estimate various measures.
Thus, approximating the size of a set using subset samples corresponds to approximating
 the support size using conditional sampling in the special case where the distribution is uniform over a subset $S$.

\subsection{Further Research}
We suggest several questions and directions for further research.
\BE
\item We showed that in the case of interval subsets, adaptive subset sampling
significantly improves the complexity of set size approximation: The dependence on $w$
is reduced from linear to polylogarithmic and the (logarithmic) dependence on $n$
is removed. The question is whether the dependence on $w$ can be further reduced (recall
that when all subsets are allowed then there is an adaptive algorithm whose
complexity is $\tilde{O}(\loglog(w))$).
\item As already mentioned, the upper bounds on the complexity of the problem
when using subset sampling with all subsets are based on subset queries.
An interesting question is whether sampling has any more power in this setting.
Indeed, obtaining lower bounds for subset sampling, in particular for adaptive
algorithms, seems more challenging than for subset queries.
%\mnote{D: added}
A related question, which was mentioned earlier as well, is whether it is possible
to formalize and analyze the relationship between the generality of the allowed subsets
and the power of subset sampling as compared to subset queries.
% \footnote{Similar difficulties arise when trying to obtain lower bounds in the
% conditional sampling model...}
\item % As noted previously, our algorithms for interval subsets can be adapted to
% higher-dimensional grids and to the Boolean hypercube.\mnote{D: return here after
% writing for hypercube}
 In this work we considered the case of interval subsets and related families of subsets.
 We believe that there are other families of subsets that may arise naturally in different
 contexts and are worth studying. %\mnote{D: changed a bit}
 % (where algorithms generalize well\mnote{D: phrasing?} to other cases), but many other families of subsets may be of interest. \mnote{What should we keep here?}
 %At this point we have some preliminary results for the case that the subsets are restricted in terms of their size.
\EE

%\subsection{Organization}

\def\allU{
\section{Querying and Sampling From $U$ with no Access to Subsets}
\label{sec:allU}

\ifnum\conf=0
This setting is the simplest to analyze  and we discuss it first.
\fi
It is clear that subset queries (in contrast to subset sampling) provide us almost no information in this setting. We can perform a single subset query (on all of $U$), and if it returns $1$ we know that $|S|\geq 1$. Otherwise, we know that $|S| = 0$. Approximating $|S|$ beyond this is impossible. It remains to discuss sampling.
Both the upper and the lower bound are based on the probability of obtaining the same element
twice (a {\em collision\/}) as a function of the size of the set, $w$.
\BT\label{thm:allU-SuS}
\BE
\item There exists a set size approximation algorithm that is provided (only) with
samples from $U$ and with high constant probability uses $O(\sqrt{w}/\eps^2)$ samples.
Furthermore, the probability that
 the number of queries it performs is larger by a factor of $k$ than this upper bound
 decreases exponentially with $k$.
% Furthermore, for any sufficiently large integer $k$, the probability that
% the algorithm uses more than $k\cdot \sqrt{w}/\eps^2$ samples is $\exp(-\Omega(k))$.
\item Every set size approximation algorithm that  is provided only with samples
 from $U$ must use $\Omega(\sqrt{w})$ samples with constant probability
for any constant $\eps$. This lower bound holds even when the algorithm is
given an estimate $\tw$ such that $\tw/4 \leq w \leq 4\tw$.
\EE
\ET

\BPF
We begin with the lower bound. For simplicity we give the argument for $\eps=1$. This directly
implies the lower bound for any $\eps < 1$.
% and $\delta < 1/3$, and generalizes in a straightforward manner to
% any constant $\eps > 1$ (and $\delta < 1/3$).
For any given $4 \leq \tw \leq |U|/4$, consider the following
two distributions over sets $S$. The first distribution is uniform over subsets
of size $\tw/4$ and the second distribution it uniform over subsets of size $4\tw$.
% By the ``birthday paradox'' (the lower bound direction),
For both distributions, the
probability that a sample of size $s= \sqrt{\tw/c}$ contains some repeated
element (a collision) is at most ${s\choose 2}\cdot \frac{4}{\tw}$
(the term ${s\choose 2}$ is
the number of sample pairs among $s$ samples,
and the term $4/\tw$ is the probability that a specific
pair is a collision in the first distribution (and an upper bound on this probability
in the second distribution)).
For $c \geq 6$ this probability is at most $1/3$.
Conditioned on there being no collisions, the samples in both cases are identically
distributed.

Turning to the upper bound, the algorithm works in two stages. In the first stage it
 obtains
an estimate $\tw$ such that $\tw/c \leq w \leq c\cdot \tw$
with probability at least $5/6$
for some constant $c> 1$ (e.g., $c=100$ suffices).
In the second stage it takes a sample of size $\Theta(\sqrt{\tw}/\eps^2)$
to obtain a $(1+\eps)$-factor approximation.
We note that while describing the algorithm as having two stages may seem to
imply that it is adaptive, this is not the case. Since the algorithm only receives
samples from $U$, the only decision it makes is when to stop and output
an estimate. Details follow.

To obtain the (rough) estimate $\tw$, the
algorithm takes samples from $U$ until a collision occurs.
If this first collision occurs on the $j^{\rm th}$ sample, then
the estimate $\tw$ is set to
$j^2$. The probability that $j < \sqrt{w/c}$
is upper bounded by ${\sqrt{w/c} \choose 2}\cdot \frac{1}{w} < 1/c$ (using a union
bound on the collision probability of each pair of elements).
 On the other hand, the probability that
$j > \sqrt{c\cdot w}$ is upper bounded by
$(1-\sqrt{c\cdot w}/(2w))^{\sqrt{c\cdot w}/2} < \e^{-c/4}$.
This upper bound follows by considering a partition of the sample
into two parts of equal size $\sqrt{c\cdot w}/2$. The probability
that no collision occurs in the sample is upper bounded by the probability
that no collision occurs in the first half of the sample, which is upper bounded by 1,
times the probability that no collision occurs between an element selected
in the second half and an element  selected in the first half (conditioned
on the elements in the first half being distinct).

It follows that with probability at least $5/6$
% the median value among the $\tw_t$s is within a factor of
$\tw$ is within a factor of 6 from $w$
% $c$ from $w$.
Furthermore, the probability that the total number of
samples taken in the first stage exceeds $k\cdot\sqrt{w}$
decreases exponentially with $k$.
%is  at most $\exp(-k/4)$. %\mnote{D: maybe can do something more careful}
% If $\eps \geq c-1$, then we are done. Otherwise we continue
% to the second stage, where if $\eps > 1/2$ we set it to $1/2$.
Assume from this point on that indeed $w/6 \leq \tw \leq 6 w$.
Also assume that $\eps \leq 1/2$, or else set $\eps$ to $1/2$.

In the second stage the algorithm
% repeats the following estimation procedure $\Theta(\log(1/\delta))$ times and sets its output,
% $\hw$, to be the median estimate. In each repetition $t$ it
takes a sample of size
$s= \Theta(\sqrt{\tw}/\eps^2)$. % Considering any fixed  $t$,
For each $1 \leq i < j \leq s$,
let $\eta_{i,j}$ be a random variable indicating whether the $j^{\rm th}$
sample is the same as the $i^{\rm th}$. Thus $\Pr[\eta_{i,j}=1]=1/w$,
and $\Exp[\sum \eta_{i,j}] = {s\choose 2}\cdot \frac{1}{w}$.
Let $\eta = \sum \eta_{i,j}$.
We set $\hw$ to be ${s \choose 2}/\eta$.
By Chebishev's inequality,
\BEQ
\Pr\Big[\Big|\eta - \Exp[\eta]\Big| > (\eps/2)\cdot \Exp[\eta]\Big]
  < \frac{4\var[\eta]}{\eps^2\Exp[\eta]^2}\;.
\EEQ
Now, $\var[\eta]= \Exp[\eta^2] - \Exp[\eta]^2$ where
\BEQN
\Exp[\eta^2]
 &=& \sum_{i<j} \Exp[\eta_{i,j}^2] + 4\sum_{i<j<k}\Exp[\eta_{i,j}\eta_{j,k}]
    + 6\sum_{i<j<k<\ell}\Exp[\eta_{i,j}\eta_{k,\ell}] \nonumber \\
 &=& {s\choose 2}\cdot \frac{1}{w} + 4{s\choose 3}\cdot \frac{1}{w^2} +
 6{s\choose 4}\cdot \frac{1}{w^2}\;.
\EEQN

Since $\Exp[\eta]^2 = {s\choose 2}^2 \cdot \frac{1}{w^2}$, which is
lower bounded by $6{s\choose 4}\cdot \frac{1}{w^2}$, we get that
 $\var[\eta] = O(s^2/w+s^3/w^2)$. It follows that
 $4\var[\eta]/(\eps^2\Exp[\eta]^2) = O(\eps^{-2}(w/s^2 + 1/s)$. By the
choice of $s$ this is at most $1/6$ (for an appropriate constant in
the $\Theta(\cdot)$ notion for $s$).
% , and by taking the median value over the
% $\Theta(\log(1/\delta))$ repetitions of the estimation procedure, we
% get a final estimate as desired.
Therefore, with probability at least $2/3$ we get that $\hw \leq (1+\eps/2)w$
and $\hw \geq (1-\eps/2)w \geq w/(1+\eps)$, as required. As for the sample size,
%the probability that it exceeds $k\cdot \sqrt{w}/\eps^2$ (due to $\tw$
%overestimating $w$), decreases exponentially with $k$.
it exceeds $k\cdot \sqrt{w}/\eps^2$ only due to $\tw$
overestimating $w$. The probability of this event (as described above) decreases exponentially with $k$.
\EPF
}
% end of allU section

\ifnum\conf=0
\allU
\fi

%%%%%%%%%%%%%%%%%%%%%%%%

\section{Interval Subset Queries and Subset Samples}\label{sec:int}
\ifnum\conf=1
Since the main focus of this work is on Interval Subsets (and related families of subsets), we
start with our algorithms and lower bounds for this family of subsets.
\fi

%Single theorem for interval group queries.
%Lower bound for adaptive min(max(log(n),w), n/w)
%upper bound is min(n/w, w*log(n))
%In Sampling intervals mention that UB holds for general families with some properties in the beginning, then claim again in the end, referring reader to the proof for intervals.

As explained in the introduction, in the case of interval subsets, the domain is a fully order set
$U = \{u_1,u_2\dots,u_n\}$, and
the subsets that can be queried/sampled, correspond to intervals of the form $\{u_i,u_{i+1},\dots,u_j\}$.
As we discuss at the end of this section, we can extend our results to $d$-dimensional grids
for constant $d$, where queries correspond to $d$-dimensional sub-grids, and to the Boolean
hypercube, where queries correspond to sub-cubes.
%%%%%%%%%%%%%%%%%%%%%%%%%%
\iffalse
One natural way of restricting the samples (or queries) we perform on subsets of a universe is to think of the elements as residing  on a discretized line one after the other. A subset is specified by the minimal and maximal elements in it, and contains all the elements between them. We call this type of subset an ``Interval". More formally:

\BD
Let $U = \{u_1,\dots,u_n\}$. A subset sample (subset query) is an {\bf Interval sample} (interval query) if the elements in the subset have consecutive indices, that is, the sample (query) is on the elements $u_i,u_{i+1},\dots,u_j$ for some values $i,j$.
\ED
%
Obviously, what subsets are considered intervals depends on a fixed ordering of the elements of $U$.
\fi
%%%%%%%%%%%%%%%%%%%%%%%%%%

%Interval subsets are ``well-behaved" in some ways, including the following: If we are given a subset $S$ of an interval $T$ that contains more than a single element, we can divide $T$ in to two interval subsets $T_1, T_2$ such that $|S|/3 \leq |S\cap T_1| \leq 2|S|/3$. This property (with any constants instead of $1/3, 2/3$) is a sufficient condition for the algorithm that uses interval subset sampling, and it therefore can be used on many more families of subsets. Such families of subsets include grids of constant dimension and the boolean lattice with its sub-lattices.

\paragraph{On the structure of our lower bounds.}
Before stating precisely and proving our results for interval subsets, we
shortly introduce an approach that we take in several of our lower bound proofs.
%\mnote{D: Also use in other subsections}
% To try and lower bound the number of queries or samples a set size approximation algorithm must perform in some setting we typically use the following approach.
We first assume towards a contradiction that there exists an algorithm that approximates the set size with constant success probability using a certain number of queries $q \leq f(n,w)$ (where $n$ is the universe size and $w$ is the size of the hidden set $S$).
% \footnote{Of course, using standard amplification techniques, this implies there exists an approximation algorithm that succeeds with any constant probability of our choice using a number of queries that is also in $o(f(n,w))$}.
We then define a distribution over pairs of sets $S_1, S_2$ such that $|S_1| < c|S_2|$ for some constant $c >0$. Clearly, for every choice of $S_1$ and $S_2$ respecting the above conditions, the algorithm we assume exists must be able to distinguish between the hidden set $S_1$ and the hidden set $S_2$ with constant probability bounded away from $1/2$ (where the probability is over the coin tosses of the algorithm and in the case of subsets sampling also over the choice of the sampled points).
However, from an averaging argument this means that there exists a deterministic algorithm that % (performs no coin tosses and)
can distinguish with a probability bounded away from $1/2$ between pairs selected from the aforementioned distribution (where the probability is over the choice of the
pair $(S_1,S_2)$ and in the case of subsets sampling also over the choice of the sampled points).
% In particular, if the algorithm is non-adaptive this means that there is a fixed set of
% $q$ queries (or samples) that distinguishes with probability bounded away from $1/2$
% between the smaller and the larger set selected by our distribution.
% To show this leads to a contradiction we then show that any algorithm
% that does not perform coin-tosses, with high (non-constant) probability
% will not distinguish between $S_1$ and $S_2$. Here the probability is
% over the distribution of pairs and over the results of samples.
Whenever the algorithm performs a subset query or asks for a subset sample from one of the sets,
we give it ``for free'' the result of the query/subset sample on the other set as well.
To obtain the lower bound we show that if the algorithm performs fewer than $f(n,w)$ queries (subsets samples),
then the answers it gets are equally/similarly distributed, implying that the algorithm
cannot distinguish between the two sets with constant probability.
% For simplicity, and as we are concerned with asymptotic results, we assume that any
% query is performed both on the universe with $S_1$ in it and on the universe with $S_2$ in it.

%Any non-adaptive algorithm that approximates set size with probability $5/6$ using $o(n/w)$ interval queries will, in particular, approximate set size for any member of a particular distribution of hidden subsets with probability $5/6$. Thus, if we give two distributions, one with with hidden subsets of size $\hw$ and the other with hidden subsets of size $2\hw$, from an averaging argument there will exist a set of interval group queries of size $o(n/w)$ that will allow an algorithm to distinguish between members of the light distribution and the heavy one. We will show two distributions that have the

\subsection{Interval Subset Queries}

\BT\label{thm.interval.group}
The following holds for the query complexity of set size approximation algorithms that use only interval subset queries:
\BE
\item\label{interval.group.nonadaptive.UB}
There exists a non-adaptive set size approximation algorithm that performs $O(n/(w\epsilon^2))$ interval queries with high constant probability. Furthermore, the probability that
 the number of queries it performs is larger by a factor of $k$ than this upper bound
 decreases exponentially with $k$.
\item\label{interval.group.nonadaptive.LB}
 Any non-adaptive set size approximation algorithm that performs only interval queries performs $\Omega(n/w)$ such queries with constant probability (for constant $\eps$).
 This lower bound holds even when the algorithm is given a constant factor
 approximation, $\tw$  of $w$.
\item\label{interval.group.adaptive.UB} There exists an adaptive set size approximation algorithm that always performs   $O(w\log(n))$ interval queries and with high constant probability performs $O(n/(w\eps^2))$ interval queries. Furthermore, the probability that
 the number of queries it performs is larger by a factor of $k$ than the latter upper bound
 decreases exponentially with $k$.
\item\label{interval.group.adaptive.LB} Any adaptive set size approximation algorithm performs $\Omega(\min(w\log(n/w^2), n/w))$ interval queries with constant probability (for constant $\eps$). This lower bound holds even when the algorithm is given a constant factor
 approximation, $\tw$ of $w$.
\EE
\ET

\BPF
Verifying all the items in Theorem~\ref{thm.interval.group} aside of Item~\ref{interval.group.adaptive.LB} is fairly simple. We begin by explaining them briefly and then turn to the proof of Item~\ref{interval.group.adaptive.LB}.

\paragraph{Item~\ref{interval.group.nonadaptive.UB}.}
The algorithm referred to in this item uses the fact that single elements are a special case of intervals. The algorithm works in two stages. In the first
 stage it obtains a rough estimate $\tw$ of $w=|S|$. This is done by iteratively selecting
 single elements
uniformly at random from $U$ and performing subset queries on them. This stage ends once
the algorithm receives a positive answer in some iteration $j$, and $\tw$ is set to
$n/j$. For any $c>1$, the probability that $j < n/(c\cdot w)$, so that $\tw > c\cdot w$
is upper bounded by $(w/n)\cdot n/(c\cdot w) = 1/c$, and the probability that
$j > c\cdot(n/w)$, so that
$\tw < w/c$ is upper bounded by $(1-w/n)^{c\cdot(n/w)} < e^{-c}$.
In the second stage of the algorithm it selects
$s = \Theta(n/(\tw\eps^2))$
elements uniformly at random, makes a subset query on each, and sets its output, $\hw$
to be $(n/s)$ times the fraction of queries that were answered positively.
By the multiplicative Chernoff bound, conditioned on $\tw \leq 6 \cdot w$ (so
that $n/\tw \geq n/(6w)$), the estimate $\hw$, is as required with probability
at least $5/6$. Since the probability that $\tw \leq 6 \cdot w$ is at most $1/6$,
we get that $\hw$ is as required with probability at least $2/3$. The furthermore claim
in this item directly follows from the
aforementioned bound on the probability that $\tw$ underestimates $w$.

\paragraph{Item~\ref{interval.group.nonadaptive.LB}.}
For the lower bound in this item consider partitioning the universe into intervals of size $\tw$, denoted $I_{1},\dots,I_{n/\tw}$. We select a value $j^\ast$ uniformly at random from
$\{1,\dots,n/\tw-1\}$. The smaller set $S_1$ consists of all the elements in $I_{j^\ast}$. The larger set $S_2$ consists of all the elements in $I_{j^\ast}\cup I_{j^\ast+1}$.
Thus $|S_1| = \tw$ and $|S_2| = 2\tw$.
Consider any sequence of interval queries $T_1,\dots,T_q$ where $q < n/(12\cdot \tw)$.
The probability (over the choice of $j^\ast$), that
for some set $T_i$, at least one of its endpoints belongs to $I_{j^\ast}$ or $I_{j^\ast+1}$
is at most $1/3$. But if such an event does not occur, then the answers to the
queries $T_1,\dots,T_q$ are the same for $S_1$ and $S_2$ and they cannot be distinguished.
% Obviously only a set of interval queries where at least one end of one of the queried subsets falls within the subinterval $I_{i+1}$ will  distinguish between these two sets. Thus, any fixed set of queries of size $o(n/w)$\mnote{D: $o(\cdot)$ or $n/(cw)$?} cannot distinguish with constant probability between $S_1$ and $S_2$. The lower bound follows.

\paragraph{Item~\ref{interval.group.adaptive.UB}.}
The algorithm referred to in this item combines two procedures. The first procedure,
 which performs $O(w\log(n))$ queries (deterministically),
determines the set $S$ exactly. %\mnote{D: elaborated a bit}
This is done by performing a type of ``extended'' binary search.
That is, the search constructs a binary tree, where each node in the tree corresponds to
an interval query. The root of the tree corresponds to all of $U$. If a query on an interval
$I$ is answered positively where $|I| > 1$, then the corresponding node has
two children, one for the left half of $I$ and one for the right. A node is a leaf if
  either the answer on the corresponding interval is negative, or if the answer is positive and
  the interval is of size $1$. Thus, we have a non-empty leaf for every element of $S$, and
  the total number of queries performed is $O(w\log n)$.
  The second procedure runs the non-adaptive algorithm of Item~\ref{interval.group.nonadaptive.UB} to estimate the size of $S$. By performing interleaved queries and stopping when the first procedure stops the desired bound is obtained.

\paragraph{Item~\ref{interval.group.adaptive.LB}.}
For this lower bound we first address the (simpler) case in which
$\tw \geq \sqrt{n}$ (so that the lower bound should be $\Omega(n/w)$).
The construction is the same as in Item~\ref{interval.group.nonadaptive.LB}, except that here
$S_1$ and $S_2$ also include the first element from each interval $I_j$.
Since the number of intervals is $n/\tw \leq \tw$, we have that
% $\tw \leq |S_1|\leq 2\tw$, $2\tw \leq |S_2| \leq 3\tw$,
% and $|S_2| \geq (3/2)|S_1|$.
$|S_1|$ and $|S_2|$ are both $\Theta(\tw)$ and $|S_2|$ is a constant factor larger
than $|S_1|$.
By this construction, any query that
does not contain {\em only\/} elements of a single interval $I_j$, is answered
positively both  for $S_1$ and for $S_2$. On the other hand, as long as the
algorithm does not ask a query $T\subset I_j$ for $j \in \{j^\ast,j^\ast+1\}$, it will get
a negative answers for both $S_1$ and $S_2$. The lower bound follows.
% Consider partitioning the universe into $n/\tw \leq \tw$ sub-intervals of size $\tw$ each.
% The first element of each of these subintervals will be a member of any set $S_1$
% and any set $S_2$ we choose - indeed the set $S_1$ in all the pairs will
% contain exactly these elements. The set $S_2$ will also contain all the
% elements in one of these subintervals, where the ``special'' subinterval
% is selected uniformly at random. By this construction, any query that
% does not contain {\em only\/} elements of a single subinterval
% Obviously any distinction between these two sets requires $\Omega(n/\tw)$ queries,
% as any query which does not contain {\em only\/} elements of a single subinterval
% will be answered by $1$ by both sets
% and will give us no information, and we must query the ``correct"
% subinterval to distinguish between $S_1$ and $S_2$.

It remains to address the case that $\tw < \sqrt{n}$.
To gain intuition, consider the following ``game''. There are $b$ identically looking
locked boxes where only one of these boxes is non-empty. We would like to open the non-empty
box, but opening each box requires time $t$, and until it is open we do not know whether
it is empty or not. If the non-empty box is selected uniformly at random, then, with high
constant probability, it will take us time $\Omega(bt)$ to get to the non-empty box.

To obtain a lower bound following the above intuition, we partition $U$ into $\tw$ intervals $I_1,\dots,I_{\tw}$ of size $n/\tw$ each (each of these corresponds to a ``locked box'').
Similarly to the construction for $\tw \geq \sqrt{n}$,
 both $S_1$ and $S_2$ include the first element from each interval. In addition,
we further partition each interval $I_j$ into $m = \lceil n/\tw^2\rceil $ subintervals
$I_j^1,\dots,I_j^m$ of size $\tw$ each (more precisely, of size at least $\tw/2$ and at most
$\tw$).
% \mnote{D: may need smaller for $\tw$ close to
% $\sqrt{n}$ or set threshold between case differently}.
For both $S_1$ and $S_2$, we select, uniformly at random, one subinterval
%\mnote{D: disallow first subinterval? }
$I_j^{\ell(j)}$ from each interval $j$, and both sets include the first and the
last element from each of these subintervals. In addition, we select a ``special'' interval
$I_{j^\ast}$ uniformly at random, and include in $S_2$ {\em all\/} elements in the
subinterval $I_{j^\ast}^{\ell(j^\ast)}$. Hence, $|S_1|$ and $|S_2|$ are both
$\Theta(\tw)$ and $|S_2|$ is a constant factor larger than $|S_1|$.

Consider any (possibly adaptive) deterministic algorithm. We would like to show that
in order to distinguish correctly between $S_1$ and $S_2$ with probability at least $2/3$
(over the choice of $S_1$ and $S_2$), it must perform at least
$\tw\cdot\log(n/\tw^2)/6$ queries with constant probability.
Consider a {\em process\/}
that answers queries of an algorithm while it selects the pair $(S_1,S_2)$ on the fly
(according to the aforementioned distribution). Actually, the process
makes all the random choices for $S_1$ and $S_2$, except the choice of $j^\ast$. That is,
it selects $\ell(j)$ for each $j$ in advance, but does not select $j^\ast$.

For each interval query $T$, if $T$ is not
strictly contained within one of the intervals $I_j$, then the process answers positively
for both $S_1$ and $S_2$. For each
interval query $T$ that is  contained within an interval $I_j$, if
 $T$ is not  contained within the subinterval $I_j^{\ell(j)}$,
then
the process gives the same answer for both $S_1$ and $S_2$ (i.e., a positive
answer if $I_j^{\ell(j)} \cap T \neq \emptyset$
and a negative answer if $I_j^{\ell(j)} \cap T=\emptyset$).
Once the algorithm performs a query $T$
% with at least one endpoint\mnote{G: Don't we need both endpoints for it to be different between $S_1,S_2$. Elaborate query process}
that is  contained in $I_j^{\ell(j)}$,
which we'll refer to as {\em revealing $\ell(j)$\/},
then the process does the following. Let $r$ be the number
of indices $j$ for which the algorithm has not yet revealed $\ell(j)$
(where initially $r = \tw$). With probability $1/r$ the algorithm sets $j^\ast = j$
(possibly implying a different answer for $S_1$ and $S_2$, so that the algorithm ``wins'').
 Otherwise it
increases $r$ by $1$ and gives the same answer for both sets (positive if $T$
contains at least one of the end-points of $I_j^{\ell(j)}$ and negative otherwise).

It follows that in order to distinguish between $S_1$ and $S_2$ with probability
at least $2/3$, the algorithm must reveal at least a one-third of the $\ell(j)$s
with probability at least $1/3$.
% DETAILS: otherwise, it can succeed with probability at most $1/3\cdot 1 + 2/3\cdot (1/2)$
% where the first time is an upper bound on the probability that it succeeds when revealing
% at least 1/3, and the second terms is the probability that it success when revealing less.
It remains to show that such a task requires at least
$\tw\cdot \log(n/\tw^2)/6$ queries with constant probability. In order to obtain
this lower bound, we consider a closely related problem, which we refer to as
{\em multiple single elements\/} that we define next. Each instance of this problem
is parameterized by two integer parameters, $b$ and $m$. It consists of $b$
binary strings of length $m$ each, where in each string $s^j$, there is a single
1 in a position $\ell(j)$, and all other bits are 0. An algorithm may ask, for
any substring $t$ that is a (consecutive) substring of some $s^j$ whether it is
all 0 or contains a 1. The goal of the algorithm is to to determine ({\em reveal\/})
$\ell(j)$ for at least $b/3$ of the substrings $s^j$. We next prove that
any algorithm for the {\em multiple single elements\/} problem, when given as
input a uniformly selected random instance (i.e., in which each $\ell(j)$ is selected
uniformly at random in $\{1,\dots,m\}$), must perform at least
$b\cdot\log(m)/c$ queries with high constant probability (for a sufficiently large
constant $c$). This will imply the lower bound claimed in this item, since
an algorithm that distinguishes between $S_1$ and $S_2$ with probability at least $2/3$
can be used for this problem.\footnote{The (straightforward) reduction is achieved by mapping
between elements of each string $s^j$ and the subintervals $I_j^\ell$ of $I_j$.}

For any algorithm that solves the multiple single elements problem, consider the decision
tree that corresponds to it. That is,  each internal node in the tree corresponds to
a query, and each leaf corresponds to an output $\ell(j_1),\dots,\ell(j_{b/3})$.
The number of leaves at depth at most $b\cdot\log(m)/6$ is $m^{b/6}$. For any
fixed leaf, the fraction of instances for which it contains a correct answer  is $m^{-b/3}$.
Therefore, the total fraction of instances for which some leaf at depth
at most $b\cdot\log(m)/6$ provides a correct answer is $m^{-b/6}$,
and the lower bound follows.
\EPF

\subsection{Interval Subset Samples}

\BT
The following holds for the sample complexity of set size approximation algorithms that use only interval subset samples:
\BE
\item\label{interval.sampling.nonadaptive.LB} Any non-adaptive set size approximation algorithm that performs only interval samples requires $\Omega(\min(n/w,\sqrt{w}))$ such samples.
\item\label{interval.sampling.nonadaptive.UB} There exists a non-adaptive set size approximation algorithm that uses $O(\min(n/w,\sqrt{w})/\epsilon^2)$ interval samples. The probability of using more samples decreases exponentially with the number of samples.
\item\label{interval.sampling.adaptive.UB} There exists an adaptive set size approximation algorithm that uses $\tilde{O}(\log(w)^4/\epsilon^2)$ interval samples. The probability of using more samples decreases exponentially with the number of samples.
\EE
\ET
We note that the $\log^4(w)$ factor in Item~\ref{interval.sampling.adaptive.UB} can
be reduced to $\log^{3+\gamma}$ for any constant $\gamma$, but for the sake of simplicity,
we give the slightly higher upper bound.

\BPF
The lower bound argument for Item~\ref{interval.sampling.nonadaptive.LB} is similar to that used for Item~\ref{interval.group.nonadaptive.LB} in Theorem~\ref{thm.interval.group}. Consider partitioning the universe into intervals of size $\tw$, denoted  $I_{1},\dots,I_{n/\tw}$. We select a $j^\ast$ uniformly
at random from  $\{1,\dots,n/\tw\}$. The small set $S_1$ is composed of $\tw/2$ elements selected uniformly at random from $I_{j^\ast}$. The large set $S_2$ consists of all the elements in $I_{j^\ast}$. Consider any fixed choice of intervals $T_1,\dots,T_q$ for
$q < \min(n/\tw,\sqrt{\tw})/c$ for a sufficiently large constant $c$.
The probability, over the choice of $j^\ast$, that any one of these intervals
has at least one end-point in $I_{j^\ast}$ is at most $1/6$. If this event does
not occur, then each interval either contains $I_{j^\ast}$ or is completely
disjoint from it. In the latter case, no sample is returned for both $S_1$ and $S_2$.
In the former case, as long as a collision (repetition of an element) does not occur,
for both $S_1$ and $S_2$, each new sampled element is uniformly distributed in $I_{j^\ast}$.
By the upper bound on $q$, the probability that a collision occurs is a small constant.
% As the minimal and maximal elements (per sample) in any fixed set of samples of size $o(n/w)$ does not intersect (with high probability) the subinterval $I_{i}$, when performing $o(n/w)$ samples for both sets we get uniformly selected elements from the interval $I_i$. This holds until $\Omega(\sqrt{\hw})$ such elements are examined, where we may distinguish between the two cases.

The correctness of Item~\ref{interval.sampling.nonadaptive.UB} is based on running two (subset query) algorithms in parallel, alternating between them in the choice of intervals. The algorithm that performs $O((n/w)\cdot \eps^{-2})$ queries is described in Theorem~\ref{thm.interval.group}, Item~\ref{interval.group.nonadaptive.UB}. The algorithm that performs $O(\sqrt{w}\cdot \eps^{-2})$ queries is described in Theorem~\ref{thm:allU-SuS}. When either algorithm returns a result we return that as our result. The correctness and bounds on the number of samples follow directly from those in the descriptions of the algorithms.

We now turn to describing the algorithm referred to by Item~\ref{interval.sampling.adaptive.UB}. The basic approach in this algorithm is as follows. The algorithm constructs a sequence
of intervals $I_0,\dots,I_t$, where $I_0 = U$, $I_j \subset I_{j-1}$ for each
$j \geq 1$ and with high constant probability, $|S \cap I_t| = 1$.
Let $b_j \eqdef |S \cap I_{j-1}|/|S\cap I_{j}|$ and observe that
$|S| = |S\cap I_0| = |S \cap I_t|\cdot \prod_{j=1}^t b_j$.
For each pair of intervals $I_j$ and $I_{j-1}$, the algorithm maintains
an estimate $\widehat{b}_j$ of $b_j$ such that
  $1/(1+\eps_j) \leq \widehat{b}_j /b_j \leq 1+\eps_j$ for a sufficiently small $\eps_j$,
  and with sufficiently high probability  $1-\delta_j$.
  The output of the algorithm is $\prod_{j=1}^t \widehat{b}_j$.
  The error parameter $\eps_j$ and
  the confidence parameter $\delta_j$ should be such that
  $\prod_{j=1}^t (1+\eps_j) \leq (1+\eps)$ and $\sum_{j=1}^t \delta_j$ is a small constant.

Given a pair of intervals $I_j \subset I_{j-1}$, if $1/4 \leq 1/b_j \leq 3/4$,
then, it follows from the multiplicative Chernoff bound that
 an estimate $\widehat{b}_j$ as described above can be obtained by
asking $\ln(3/\delta_j)/\eps_j^2$ subset samples on the subset $I_{j-1}$.
Therefore, it remains to explain how each $I_j$ is selected based on $I_{j-1}$
so as to ensure (with probability at least $1-\delta_j$) that indeed
$1/b_j$ is as desired. This is done by performing $s_j = 4\ln(3/\delta_j)$ subset samples on the subset $I_{j-1}$, and ordering the selected elements $v_1\leq v_2\leq \dots\leq v_{s_j}$.
If they are all equal then the algorithm sets $t= j-1$.
Otherwise, if $I_{j-1} = [w_{j-1},w'_{j-1}]$, then $I_j = [w_{j-1},v_{s_j/2}]$.
The probability that either the algorithm terminated with $I_t$ such that
$|I_t\cap S| > 1$ or that
$b_j < 1/4$ or $b_j > 3/4$ is at most $\delta_j$.

If we set $\delta_j = 1/(10j^2)$ and $\eps_j = \eps/(100j^{3/2})$
then with probability at least $2/3$ the algorithm terminates after
$c\log(w)$ iterations (for a constant $c> 1$) with an estimate as required.\footnote{The
 setting of $\delta_j$ ensures that the sum over all $\delta_j$, which is the failure
 probability of the algorithm, converges to a constant. The setting of $\eps_j$
 ensures that the product over all $j$ of $(1+\eps_j)$ is upper bounded by $(1+\eps_j)$.}
 In such a case,
the total number of subset samples used is $c' \log(w)^4\loglog(w)/\eps^2$ for a constant
$c'$. Since the probability that the algorithm does not terminate in
$c\log(w)$ iterations (since it does not obtain an interval of size 1) is at most $1/3$,
the probability that it does not terminate in $k\cdot c\log(w)$ iterations is
$\exp(-k)$, as desired.
\EPF

\subsection{Other Related Families of Subsets}
In this subsection we describe how to modify our most efficient algorithm, which uses interval subsets
(i.e., the adaptive subset sampling algorithm), to two additional setting, thus obtaining
algorithms with complexity $\poly(\log(w))$ in these settings as well. The non-adaptive algorithms using interval subsets work as is for the above two
universes and families of subsets (since these algorithms are based on
having access to singleton subsets, and to all the universe, which also holds in these settings).
The adaptive algorithm using interval queries can be easily modified to construct a
``search-tree'' whose non-empty leaves contain single elements of the set $S$.

\paragraph{$d$-dimensional grids and sub-grid subsets.}
Let $U$ be a hypergrid $\{1,\dots,k\}^d$ where
$d$ is a constant,
and let the family of subsets that the algorithm can query/sample, correspond to
$d$-dimensional sub-grids. Thus, $n= k^d$, and the special case of $d=1$ corresponds to
interval subsets over a fully ordered universe $U$.
The modified (adaptive subset-sampling) algorithm defines a sequence of sub-grids, $R_0,\dots,R_t$ where
$R_0 = U$,  $R_{j+1} \subset R_{j}$ for every $j \geq 0$ and
$|R_t\cap S|=1$ (with high constant probability).
The main observation is that for each $j \geq 0$, there exists
a sub-grid $R_{j+1}$ such that
$|R_j \cap S|/(2 d) \leq |R_{j+1}\cap S| \leq (1-1/(2d))|R_j \cap S| $,
and that a sub-grid with similar properties can be found efficiently (with
sufficiently high probability) by sampling. For simplicity, we first
establish this observation for $d=2$, and later explain how it generalizes to larger
values of $d$.

Given a two-dimensional sub-grid $R_j$ whose lower-left corner is
$(x_j^{\min},y_j^{\min})$ and whose upper-right corner is
$(x_j^{\max},y_j^{\max})$, consider all sub-grids defined
by ``cutting'' $R_j$ along the $x$-axis. That is, sub-grids
defined by $(x_j^{\min},y_j^{\min})$ and $(x_j^{\min}+b,y_j^{\max})$
or by $(x_j^{\min}+b,y_j^{\min})$ and $(x_j^{\max},y_j^{\max})$
(for $1 \leq b \leq x_j^{\max}-x_j^{\min}-1$). If one of these
sub-grids contains at least one-fourth of the points in $R_j \cap S$,
then we are done. Otherwise, there must
be a value $b^\ast$ such that the one-dimensional
sub-grid defined by $(x_j^{\min}+b^\ast,y_j^{\min})$
and $(x_j^{\min}+b^\ast,y_j^{\max})$ contains at least half of the
points in  $R_j \cap S$. But this implies that there exists
a sub-grid defined by $(x_j^{\min},y_j^{\min})$ and
$(x_j^{\max},y_j^{\min}+b')$ that contains  at least
one-fourth of the points in $R_j \cap S$.

The algorithm asks for $\Theta(\log(1/\delta_j))$ subsets samples with the subset
 $R_j$. By the above discussion and a multiplicative Chernoff bound,
 with probability at least $1-\delta_j$ there is
 either a sub-grid defined by $(x_j^{\min},y_j^{\min})$ and $(x_j^{\min}+\hat{b},y_j^{\max})$
 that contains between one-eighth and two-eights of the sample points, or there exists
 such a sub-grid defined by $(x_j^{\min},y_j^{\min})$ and $(x_j^{\max},y_j^{\min}+\tilde{b})$.
 The algorithm lets $R_{j+1}$ be this sub-grid. As in the case of interval subsets,
 the algorithm estimates the ratio between $|R_{j+1}\cap S|$ and $|R_j\cap S|$
 and uses the estimates to compute its final output.

The argument generalizes
to $d > 2$ by applying an iterative process that considers the
$d$ dimensions one after the other, ``losing'' at most a fraction
of $1/d$ of the points in each iteration. In order to ensure that
with high probability the algorithm selects a ``good'' sub-grid in
each iteration, the setting of $\delta_j$ should be reduced by a factor
of $d$, and the size of the sample required to estimate the ratio
between $|R_{j+1}\cap S|$ and $|R_j\cap S|$ needs to be increased
by another factor of $d$ (as we need to take a union bound over
the $d$ dimensions). Since $d$ is assumed to be a constant, the
complexity remains as in the case of interval subsets.

\paragraph{The Boolean hypercube and sub-cube subsets.}
Let $U = \bitset^d$ for $d = \log n$ and let the family of allowed subsets consist
of all sub-cubes (i.e., subsets of $U$ that are determined by restricting a subset
of the coordinates to a fixed value in $\{0,1\}$).
The first basic observation here is that given a sub-cube $C$ such that
$|C\cap S| \geq 2$, it contains
a sub-cube $C^\ast$ such that $|C\cap S|/3 \leq |C^\ast\cap S| \leq 2|C\cap S|/3$.
To verify this, assume, without loss of generality, that the restricted
coordinates of $C$ are $\{1,\dots,j\}$. Starting from $j+1$, we
restrict the unrestricted coordinates, where we always select the
restricted value for which the intersection with $S$ is larger. We
stop once we obtain a sub-cube $C^\ast$ as specified (where we must reach such
a stopping condition based on the restriction procedure).
% We refer to $C^\ast$ as being {\em useful with respect to $S$\/}.
In addition, let $C^+$ be the minimal sub-cube
that satisfies $C^\ast \subset C^+ \subseteq C$
and such that $|C^+ \cap S| \geq 5|C\cap S|/6$.

The second observation is the following. Suppose we ask for $\Theta(\log(1/\delta))$
subset samples with the subset $C$, and consider the  {\em maximal\/} sub-cubes $C'$
of $C$ that are defined by restricting coordinates $1,\dots,t$
for some $t \geq 1$ and that contain between $1/4$ and $3/4$ of the sample points. With probability
at least $1-\delta$, one of these sub-cubes will contain $C^\ast$ and be
strictly contained in $C^+$.
Furthermore, by the maximality of these sub-cubes and the condition on the number of
sample points that they contain, there are at most two such sub-cubes.
% they are disjoint, so that there are at most 4 of them.
If there is just one, then we are done. Otherwise,
we ask for
$\Theta(\log(1/\delta))$ additional subset samples with the subset $C$,
and among the two sub-cubes, select one that contains between
$1/4$ and $7/8$ of the sample points. In this manner we can obtain
(with high constant success probability) a sequence of sub-cubes
$C_0 \supset C_1 \supset \dots \supset C_t$ where $C_0=U$,
$|C_t\cap S| = 1$, and $|C_{j+1}\cap S|$ is a constant fraction of
$|C_j\cap S|$. By estimating these fractions, we obtain
an estimate of $S$, similarly to the case of intervals.

\ifnum\conf=1
\allU
\fi

\section{Unrestricted Subset Queries and Subset Samples}\label{sec:unrestricted}
In this section we prove the next two theorems, and get the corollary that follows. As mentioned in the introduction, analogues of these results with regard to the adaptive case were proved by Stockmeyer~\cite{Stockmeyer85}, who did not relate the complexity to $w$ or $\epsilon$ but rather only to $n$. We provide all details for the sake of
consistency.

\BT\label{thm:arbitrary-ub}
\BE
\item\label{it:arbitrary-ub-na} There exists a non-adaptive set size  approximation algorithm that performs
$\tilde{O}(\log(w)/\eps^3)$ (unrestricted) subset queries with high constant probability. Moreover,
for any integer $k$, the probability that the algorithm performs a number of queries
 that is more than a factor of $k$ larger than the above
upper bound decreases exponentially with $k$.
\item\label{it:arbitrary-ub-a} There exists an adaptive set size approximation algorithm that performs
$\tilde{O}(\loglog(w)/\eps^3)$ queries with high constant probability. Moreover,
for any integer $k$, the probability that the algorithm performs a number of queries
 that is more than a factor of $k$ larger than the above
upper bound decreases exponentially with $k$.
\EE
\ET
Stockmeyer further proved that a much smaller family of subsets, polynomial in $n$ of size, may be used to achieve similar results.
As a directly corollary of Theorem~\ref{thm:arbitrary-ub} we get that the same
upper bounds hold when the algorithm may perform subset samples.
We comment that for constant $\eps$, Item~\ref{it:arbitrary-ub-na}
in Theorem~\ref{thm:arbitrary-ub} is implied by Item~\ref{it:arbitrary-ub-a},
since any adaptive subset-query algorithm can be emulated by a non-adaptive
subset-query algorithm at an exponential cost in the query complexity.
However, since we are interested in a polynomial dependence on $1/\eps$,
we address the two cases separately.

\BT\label{thm:arbitrary-lb}
\sloppy Every non-adaptive set size approximation algorithm must perform
%$\Omega(\log(w)/\loglog(w))$ queries with probability at least $1/6$ (for
$\tilde{\Omega}(\log(w))$ queries with probability at least $1/6$ (for
any constant $\eps$).
\ET

\BC\label{cor:arbitrary-lb.adaptive}
Every adaptive set size approximation algorithm  must perform
$\tilde{\Omega}(\loglog(w))$ queries with probability at least $1/6$ (for constant $\eps$).
\EC

In the other settings we study, the lower bounds we presented held even if the algorithm was given a constant factor approximation $\tw$ of the size of $S$. In contrast, when we are allowed to query arbitrary subsets of $U$, given such an estimate $\tw$, it is possible to obtain a $(1+\eps)$-approximation
by performing a number of queries that depends only on $1/\eps$. Hence, the issue is
essentially to find such an estimate $\tw$, and this is where the dependence on $w$ comes into play.
% we can verify and refine that guess. To do this
% Specifically, our algorithms repeatedly query subsets where every element of $U$ participates with probability $1/\tw$, and the number of subsets that contain members of $S$ allows us to verify and refine our estimation of $|S|$. Thus, the upper bounds we suggest are based on searching for such a guess $\tw$, and the lower bounds are on the ability to find such a guess.
The upper bounds we suggest are based on starting with a small estimated value $\tw=1$ and increasing the estimate. In a similar manner one could start with $\tw=n$ and iteratively decrease it, leading to an upper bound of $\tilde{O}(\log(n/w)/\eps^3)$. We could take the minimum between these two values of upper bounds, but then one may start estimating $w$ from, e.g., $\sqrt{n}$. Indeed for every relation between $w$ and $n$ there is an algorithm that performs $q(\eps)$ queries for this particular relation. % This leads to a slightly strange formulation of what the lower bound actually guarantees. \mnote{Put this better.} In both lower bounds presented in this section the statement is actually that for every approximation algorithm $A$ and for every size $w$ it holds that there exists  a value $n$ such that the number of queries in the lower-bound holds.
Therefore, in the current setting, a lower bound of $\Omega(g(w))$ means that every algorithm must
perform $\Omega(g(w))$ queries for most values of $w$.
% while in the other settings a lower bound of $g(n,w)$ meant that for every
% (sufficiently large) $n$ and
% $\tw\in \{1,\dots,n\}$, every algorithm for approximating
% the size $w$ of $S$ when $w = \Theta(\tw)$ requires $g(n,\tw)=g(n,w)$ queries (for constant $\eps$),
% here lower bounds have a different meaning. They mean that the lower bound holds
% that for every size $w$ and
% every set size approximation algorithm, there exists a value $n$ for which the
% lower bound holds.\mnote{D: I am not sure about this.. Maybe say that for every $n$ this
% bound holds for most $w$'s?}
 %the query complexity of the algorithm is $g(w)$ for infinitely many settings of $w$.

\medskip
\BPFOF{Theorem~\ref{thm:arbitrary-ub}}
Both the non-adaptive and the adaptive algorithm search for
an estimate for $w$ by working in iterations, where in each iteration they hold
a hypothesis for such an estimate and they test it.
The reason
for the exponential difference between the complexities is that in the non-adaptive case
 we perform a ``doubling search'' on the size of the set, while in the adaptive
 case we perform both a doubling search and a binary search on log the size
 of the set. We first describe the algorithms for the special case of $\eps=1$, and later
explain how to modify them to other values of $\eps$.

%\paragraph{The non-adaptive algorithm (for $\eps=1$ and $w \leq n/w$).}
\paragraph{The non-adaptive algorithm (for $\eps=1$).}
For any integer $i$ let $e_i \eqdef 2^{i-1}$, and let $\delta_i \eqdef 1/(10i^2)$.
Also define $\rho_i \eqdef (1-1/e_i)^{e_i}$ (so that for $i\geq 2$, $1/4 \leq \rho_i \leq 1/e$).
The algorithm works in iterations.
In iteration $i$, for each $1 \leq j \leq t_i=c\log(1/\delta_i)$ (where $c$ is a constant)
it selects a subset $T^j_{i}$ by including each element
from $U$ in $T^j_{i}$ independently with probability $1/e_i$
and it performs a subset query on  $T^j_{i}$.
Let $\widehat{p}_i$ denote the fraction of subsets
$T^j_i$ that return a negative answer (among the $t_i$ subsets queried).
The estimate output by the algorithm is $e_i$ for the first $i$
such that $\widehat{p}_i \in [\rho_i^{\sqrt{2}}-0.02,\rho_i^2-0.02]$.

For each $i$  % such that $e_i < w/(1+\eps)$
let $p_i(w)$ denote the probability (over the choice of each $S^j_i$) that
the query on $T^j_i$ is answered negatively. By the process of
selecting each $T^j_i$ we have that  $p_i(w) = (1-1/e_i)^w = \rho_i^{w/e_i}$.
If $e_i < w/2$, then $p_i(w) < \rho_i^{2}$,
and if $e_i > 2w$, then
$p_i(w) > \rho_i^{1/2}$. %>\sqrt{e}\cdot \rho_i$.
On the other hand, there exists an index $i(w)$
such that $w/\sqrt{2} \leq e_{i(w)} \leq \sqrt{2}w$, implying that
$\rho_{i(w)}^{\sqrt{2}} < p_{i(w)}(w) \leq \rho_{i(w)}^{1/\sqrt{2}}$.
By the setting of the $t_i$s (for a sufficietly large constant $c$), for each (fixed choice of)
$i$, the probability that $|\widehat{p}_i-p_i(w)| > 0.02$ is at most $\delta_i$.
Summing over all $i$, the probability that such an event occurs for some $i$
is upper bounded by $1/3$.
Since $\rho_i^{\sqrt{2}}-\rho_i^2 > 0.04$ and $\rho_i^{1/2}-\rho_i^{1/\sqrt{2}} > 0.04$
for every $i$, we get that
the probability that the procedure outputs $e_i$
such that either $e_i > 2w$ or $e_i < w/2$ is at most $1/3$, as required.
In each iteration $i$ the algorithm performs $O(\log(1/\delta_i))=O(\log(i))$ queries,
and by the above analysis, the probability that it performs more than
$k\cdot c\log(w)\loglog(w)$ queries (for a fixed constant $c$ and any $k$)
decreases exponentially with $k$.
% If the algorithm stops in iteration $i^{\ast}$, outputing $e_{i^\ast}} = 2^{i^\ast-1}$
% which equals $k^\ast\cdot w$, then
% it performs $\sum_{i=1}^{i^{\ast}}\log(i)$

%\paragraph{The adaptive algorithm (for $\eps = 1$ and $w \leq n/w$).}
\paragraph{The adaptive algorithm (for $\eps = 1$).}
The adaptive algorithm works in two stages. In the first stage,
rather than increasing its ``hypothesis'' for $w$ (i.e., $e_i$) by a factor
of $2$ in each iteration, the adaptive algorithm increases the log of the estimate
(i.e., $i$) by a factor of $2$ in each iteration. Namely, for each integer
$\ell \geq 0$, in iteration $\ell$ it
tests the hypothesis that $e_{2^\ell} \geq  w$ and stops in the first iteration $\ell$
in which the test passes. Similarly to the non-adaptive case, this is done
by selecting $t_\ell = c\log(1/\delta_\ell)$ subsets $S^j_\ell$ (for
$\delta_\ell = 1/(10\ell^2)$), where
each element is included in $S^j_k$ with probability $1/e_{2^\ell}$
and performing a subset query on each subset. The fraction of subsets
on which a negative answer is returned is denoted $\widehat{p}_\ell$.
The test is said to pass if $\widehat{p}_\ell > \rho_{2^\ell}^{1/2}-0.01$.
Similarly to the analysis
of the non-adaptive case, with probability at least 5/6, the test
stops in iteration $\ell^\ast$ such that $w \leq e_{2^{\ell^\ast}} \leq w^2$.
That is, $\log(w) \leq 2^{\ell^\ast} \leq 2\log(w)$, so that
$\loglog(w) \leq \ell^\ast \leq \loglog(w)+1$.

In its second stage, the algorithm performs a binary search
for $i(w)$ (where $i(w)$ is as defined in the analysis of the non-adaptive algorithm)
between $i_{\min}=2^{\ell^\ast-1}$ and $i_{\max}= 2^{\ell^\ast}$. Each step of the
binary search computes an estimate $\widehat{p}_i$ as in the non-adaptive
algorithm until obtaining
an index $i$ such that $\widehat{p}_i \in [\rho_i^{\sqrt{2}}-0.02,\rho_i^2-0.02]$.
The analysis of the quality of the estimate is as in the non-adaptive case.

\paragraph{Dealing with $\eps < 1$.} In the non-adaptive algorithm and in the second stage of the adaptive
algorithm we set
$e_i \eqdef \lfloor (1+\eps/4)^{i-1} \rfloor$ and $t_i=c\log(1/\delta_i)/\eps^2$
(the first stage of the adaptive algorithm remains unchanged).
Since for any constant $x$ and $\gamma <1$ we have that $x^{1+\gamma}-x = \Omega(\gamma)$,
we modify the rule for the selected index $i$ (in the non-adaptive algorithm
and in the second stage of the adaptive algorithm) to be
 $\widehat{p}_i \in [\rho_i - \eps/c',\rho_i^{1+\eps/4}+\eps/c']$ (for an appropriate
 constant $c' > 1$). The analysis is adapted in a straightforward manner, and adds an additional factor of $O(1/\epsilon)$ due to the modified definition of $e_i$.
%\paragraph{Removing the promise that $w \leq n/w$.}
%If there is no promise that $w \leq n/w$, then, in addition to performing a search
%for $w$ in which each hypothesis is larger than the previous when, both algorithms
%also perform a search, starting from the hypothesis that $w=n$, in which each hypothesis
%is smaller than the previous one. To make this concrete, consider the non-adaptive algorithm
%for $\eps=1$. In each iteration $i$, in addition to selecting  sets $T^j_i$ as described
%above, it also selects sets $R^j_i$, where each element in $U$ is included in $R^j_i$
%with probability $1/e_{\log(n)-i}$. Let $\widehat{q}_i$ be the fraction of sets $R^j_i$
%on which the group query returned a negative answers. If
%$\widehat{q}_i \in [(\rho_{\log(n) - i})^{\sqrt{2}}-0.02,(\rho_{\log(n)-i})^2-0.02]$,
%then the algorithm outputs $e_{\log(n)-i}$ as its estimate. In this manner, if
%$w > n/w$, then the search reaches $e_{i(w)}$ after $\log(n/w) < \log(w)$ steps.
%When $\eps < 1$,
%the only modification in the above description is that $\log(n)$ is replaced by
%$\lceil \log_{1+\eps/4}(n)\rceil $. The adaptive algorithm is modified along the
%same lines.
\EPFOF

% \mnote{Consider merging with previous note}
As mentioned previously, this upper bound can be ``optimized" for different values of $w$ in relation to $n$. If we, e.g., assume
that $w > n/w$, then we can search for $w$ starting from the hypothesis that $w=n$, in an order where each hypothesis
is smaller than the previous one. To make this concrete, consider the non-adaptive algorithm
for $\eps=1$. In each iteration $i$ it selects sets $R^j_i$, where each element in $U$ is included in $R^j_i$
with probability $1/e_{\log(n)-i}$. Let $\widehat{q}_i$ be the fraction of sets $R^j_i$
on which the subset query returned a negative answers. If
$\widehat{q}_i \in \left[\rho_{\log(n) - i}^{\sqrt{2}}-0.02,\rho_{\log(n)-i}^2-0.02\right]$,
then the algorithm outputs $e_{\log(n)-i}$ as its estimate. In this manner, if $w > n/w$, then the search reaches $e_{i(w)}$ after $\log(n/w) < \log(w)$ steps.

\medskip
We now turn to the lower bounds.

\medskip
\BPFOF{Theorem~\ref{thm:arbitrary-lb}}
Here too we take the approach introduced at the beginning of Section~\ref{sec:int}.
% For the lower bound on the query complexity of any non-adaptive algorithm
Consider the following distribution over pairs of sets $(S_1,S_2)$.
First we select an integer $i$ uniformly at random
in $\{0,\dots,\log(n)-1\}$.
% among all even integers of size at most $\log n - 1$.
The set $S_1$ is a uniformly selected set of size $2^i$ and the set $S_2$ is a
uniformly selected set of size $2^{i+1}$. We make two simple observations.
The first is that if a query is on  a set $T$ such that
$|T| > 4i\cdot (n/2^i) $, then the probability that the answer to either $S_1$ or
$S_2$ is negative is $\exp(-\Omega(i))$. The second is that if a query is on a set $T$
such that $|T| < n/(4i\cdot 2^i)$, then the probability that the answer to either $S_1$ or
$S_2$ is positive is at most $1/(c'\cdot i)$.
 % Leave note explaining. D: not clear that need note
It follows that if the algorithm performs
less than $i/c''$ queries all of which are on sets of size greater than
$4i\cdot (n/2^i) $ or smaller than $n/(4i\cdot 2^i)$, then the probability
that the algorithm can distinguish between $S_1$ and $S_2$ is a small
constant (less than $1/6$).

Consider any fixed sequence of queries $T_1,T_2,\dots$ (where the only decision of the
(non-adaptive) algorithm
is after which query $T_q$ to stop and give its output).
% It remains to show that
% with constant probability over the choice of $i$, the prefix $T_1,\dots,T_{i/(c'\log(i))}$
% does not contain any set $T_j$ such that $n/(ci\cdot 2^i) \leq |T_j| \leq i\cdot (n/2^i)$.
We shall say that a subset $T_j$ is {\em useful\/} for $i$ if $j \leq i/(12\log(i)+24)$
and $n/(4i\cdot 2^i) \leq |T_j| \leq 4i\cdot (n/2^i)$.
The second condition is equivalent to
$\log(n) - i - (\log i + 2) \leq \log|T_j| \leq \log(n) - i +(\log i + 2)$.
Since $i < \log (n)$ so that $\log(i) < \loglog(n)$,
 a set $T_j$ can be useful for at most $2\loglog(n)+4$ values of $i$.
 Therefore, the subsets $T_1,\dots,T_q$ for $q \leq \log(n)/(12\loglog(n)+24)$
 can be useful for less than $\log(n)/6$ values of $i$. Since $i$ is selected
 uniformly at random in $\{0,\dots,\log(n)-1\}$, the probability (over the choice of $i$)
 that there is a subset $T_j$ that is useful for $i$ is at most $1/6$.
 The lower bound on any non-adaptive algorithm follows by combining this
 with the first part of the proof.
 % \mnote{D: work on this more.}
\EPFOF

\bibliography{set-size}
\ifnum\conf=0
\bibliographystyle{alpha}
\else
\bibliographystyle{plain}
\fi

\end{document}